\documentclass[review]{elsarticle}
\usepackage{graphicx}
\usepackage{lineno,hyperref}
\modulolinenumbers[5]

\journal{Journal of \LaTeX\ Templates}

\newcommand{\pp}{$pp$}
\newcommand{\pA}{$pA$}
\newcommand{\ppb}{$p$+Pb}
\newcommand{\dau}{$d$+Au}
\newcommand{\snn}{\sqrt{s_{_{NN}}}}
\newcommand{\gev}{GeV/$c$}
\newcommand\pt{p_T}

\newcommand\dphi{\Delta\phi}
\newcommand\deta{\Delta\eta}
\newcommand\mean[1]{\langle#1\rangle}

\begin{document}
\begin{frontmatter}
\title{Dihadron correlations in \dau\ collisions from STAR}
\author{Fuqiang Wang (for the STAR Collaboration)}
\ead{fqwang@purdue.edu}
\address{Physics Department, Purdue University, West Lafayette, Indiana 47907, USA}
\begin{abstract}
Dihadron correlations are reported for peripheral and central \dau\ collisions at $\snn=200$~GeV from STAR. The ZYAM background-subtracted correlated yields are larger in central than peripheral collisions. The difference is mainly caused by centrality biases to jet-like correlations. Fourier coefficients of the raw dihadron correlations are also reported. It is found that the first harmonic coefficient is approximately inversely proportional to event multiplicity, whereas the second harmonic coefficient is approximately independent of event multiplicity.
\end{abstract}
\end{frontmatter}

\section{Introduction}



Normally, \dau\ (and \pp\ and \pA) collisions are used as reference for heavy-ion collisions. For example, \dau\ data were essential in establishing jet-quenching at RHIC--that the observed high-$\pt$ suppression~\cite{Adcox:2001jp,STARsuppression130GeV} is not due to initial-state differences in parton distributions inside proton and nucleus but final-state parton-parton interactions and partonic energy loss~\cite{Adler:2003ii,STARdAu03}. This, in part, led to the paradigm of strongly interacting quark-gluon plasma~\cite{GyulassyMcLerran}.

Surprisingly, a long-range pseudorapidity ($\deta$) dihadron correlation at small azimuthal difference ($\dphi$) was observed, above a uniform background, at high $\pt$ in high-multiplicity \pp\ collisions at the LHC~\cite{CMSppRidge}; it was later observed in \ppb\ collisions at essentially all $\pt$ and multiplicity (except very low multiplicity)~\cite{CMSpPbRidge,ALICEpPbRidge,ATLASpPbRidge}. 
This long-range $\deta$ correlation is called the ``ridge,'' in analogy to the similar structure observed in heavy-ion collisions; see below. 
This motivated further studies of those small-system collisions, beyond just for their use as reference for heavy-ion collisions.

In fact, prior to the ridge observation in small systems at the LHC, dihadron correlations were extensively studied in \dau\ collisions at RHIC~\cite{STARridge09,Horner,Nattrass}. No ridge correlations were observable in \dau. The \dau\ dihadron correlations were similar to those in \pp\ collisions, although slight modifications were seen. A difference was observed by using the cumulant variable between \pp\ and \dau\ collisions~\cite{STARflow05}, qualitatively consistent with a slight difference in dihadron jet-like correlations. 

Similar ridge correlations had been observed before in heavy-ion collisions after a subtraction of elliptic anisotropy background~\cite{PRL95,STARridge09,PHOBOSridge10}. The heavy-ion ridge correlations were attributed primarily to triangular anisotropy~\cite{AlverV3}. The similarity of the ridge in \pp\ and \ppb\ collisions, where only a uniform background is subtracted, suggests that elliptic anisotropy may be responsible for the ridge in these small systems. In fact, hydrodynamic calculations with event-by-event geometry fluctuations can qualitatively and semi-quantitatively describe the observed ridge in \pp\ and \ppb\ collisions~\cite{Bozek:2010pb,Bozek:2011if,Bozek:2012gr}. Whether the experimentally measured azimuthal anisotropies are of hydrodynamic flow origin remains a quantitative open question.

Hydrodynamic flow is not the only explanation for the \pp\ and \ppb\ ridge correlations. They can also be described by the Color Glass Condensate, where two-gluon density is relatively enhanced at small $\dphi$ over a wide range of $\deta$~\cite{Dumitru:2010iy,Dusling:2012iga,Dusling:2012wy,Dusling:2013oia}.

Recently, a back-to-back double ridge, resembling an elliptic/quadrupolar shape, was observed by subtracting the per-trigger normalized dihadron correlated yield in peripheral \ppb\ collisions from that in central collisions~\cite{ALICEpPbRidge,ATLASpPbRidge}. If the correlated jet fragments--dominating the away-side dihadron correlations at large $|\deta|$--are equal in number between peripheral and central collisions, then the observed double ridge would be an indication of new physics. Jet correlations are due to hard scattering and are not expected to differ, in leading order, over \ppb\ collision centrality, except that the centrality definition, usually by measured multiplicity in the final state, can bias the selection of events with varying magnitudes of jet correlations. For example, events selected by higher multiplicity could contain jets, both near and away side--because they contribute to the overall multiplicity measurement--originating from partons with larger energy or softer fragmentation.
Such biases were estimated to be 10-20\%~\cite{ALICEpPbRidge} of the observed ridge yield above a uniform pedestal, indicating that the double ridge may indeed be due to new physics other than jets.

PHENIX analyzed \dau\ data using the same technique of ``central $-$ peripheral'' dihadron correlations in the limited $\deta$ acceptance of $|\deta|<0.7$ with the central arm detector~\cite{PHENIXdAuRidge}. They observed a double ridge in the ``central $-$ peripheral'' dihadron correlations. It is unclear how much centrality biases there are on jet correlations within the PHENIX acceptance. 

STAR, with its large acceptance, is suitable to investigate centrality biases to dihadron correlations. 
STAR has extensively studied \dau\ collisions~\cite{STARdAu03,STARridge09,Horner}.
The recent development of LHC and PHENIX data called for a more detailed study of the STAR data. This contribution reports the status of such a study.

\section{Data Sample}

The data presented here were taken during the \dau\ run in 2003~\cite{Levente09}. The coincidence of the signals from the Zero Degree Calorimeters (ZDC) and the Beam-Beam Counters (BBC) selects minimum-bias (MB) events of \dau\ collisions, corresponding to ($95\pm3$)\% of the total hadronic cross-section\cite{STARdAu03}. Events used in this analysis are required to have a primary vertex position $|z_{\rm vtx}|<30$~cm from the TPC center. A total of approximately 10 million events were used.  TPC(FTPC) tracks are required to have at least 25(5) out of maximum possible 45(10) hits and a distance of closest approach to the primary vertex within 3~cm.

\begin{figure*}
\begin{center}
\includegraphics[width=0.32\textwidth]{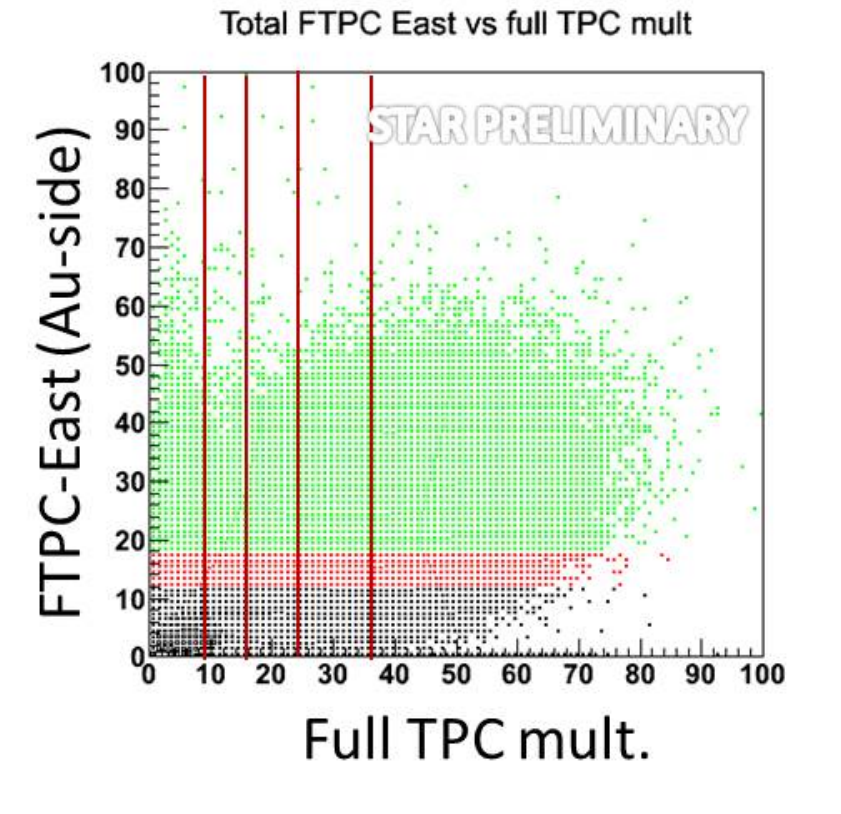}
\includegraphics[width=0.32\textwidth]{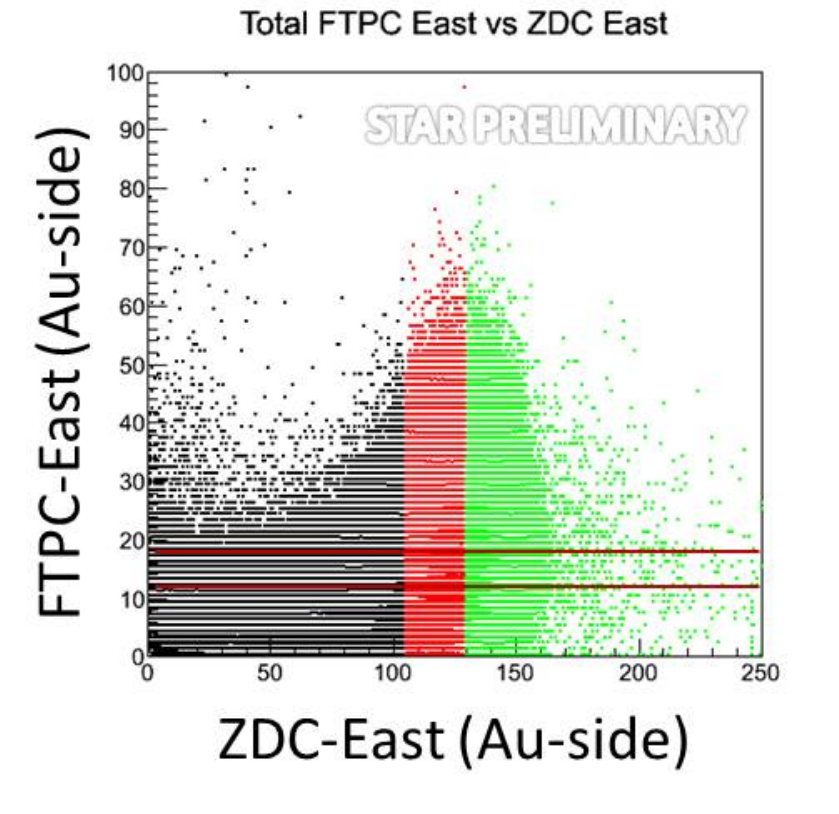}
\includegraphics[width=0.32\textwidth]{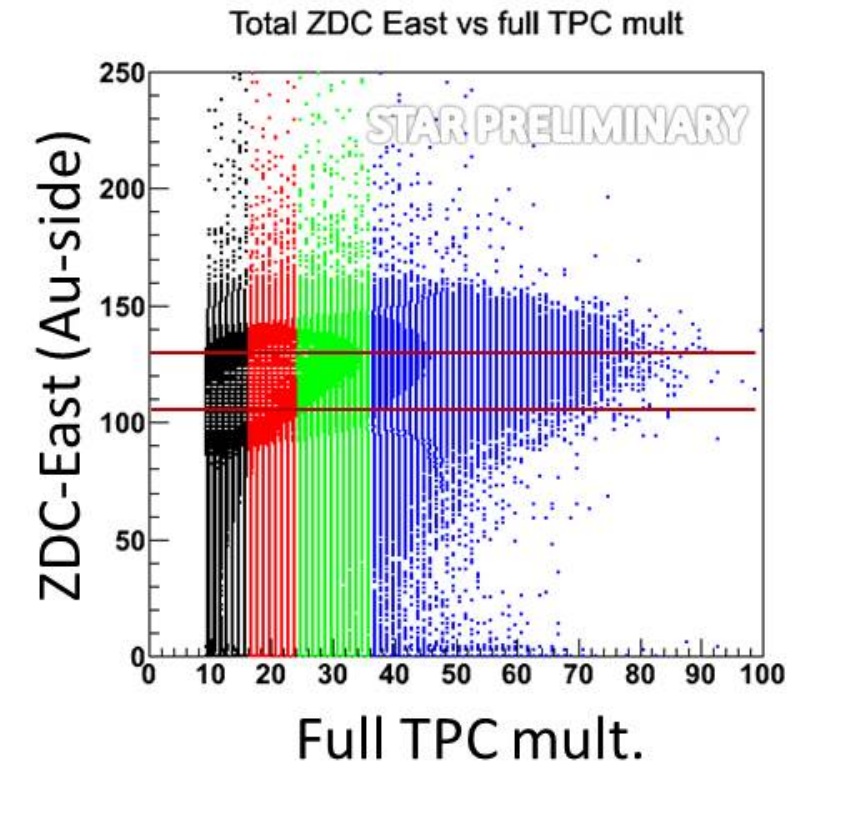}
\caption{Scatter plots of correlations between three centrality measures: TPC multiplicity ($|\eta|<1$), FTPC-Au multiplicity ($-3.8<\eta<-2.8$), and ZDC-Au neutral energy. Different colors and lines indicate the centrality ranges defined by these measures, corresponding, as examples, to 40-100\%, 20-40\%, and 0-20\% for both FTPC-Au and ZDC-Au measures, and 80-100\%, 50-80\%, 30-50\%, 10-30\%, and 0-10\% for the TPC measure.}
\label{fig:centrality}
\end{center}
\end{figure*}

Three quantities were used to define \dau\ centrality: charged particle multiplicity within $|\eta|<1$ measured by the TPC, charged particle multiplicity within $-3.8<\eta<-1.8$ measured by the FTPC in the Au-beam direction (FTPC-Au)~\cite{Levente09}, and neutral energy measured in the ZDC of the Au-beam direction (ZDC-Au). The correlations between between each pair of the three observables are shown in Fig.~\ref{fig:centrality}. Positive correlations are observed but the correlations are quite broad. The same percentile centralities defined by different centrality measures correspond to significantly different event samples of \dau\ collisions.

\section{Data Analysis}

Two sets of dihadron correlations are analyzed: TPC-TPC correlations where the trigger and associated particles are both from the TPC within $|\eta|<1$, and TPC-FTPC correlations where the trigger particle is from the TPC and the associated particle is from either the FTPC-Au within $-3.8<\eta<-2.8$ or the FPTC-d within $2.8<\eta<3.8$. The $\pt$ ranges of trigger and associated particles reported here are both $1<\pt<3$~\gev. (For TPC-TPC correlations the trigger and associated particles are, as such, the same set of particles in each event but each pair is counted only once.) The associated particle tracking efficiencies, $\epsilon_{\rm assoc}=85\%$ for TPC tracks and $70\%$ for FTPC tracks (both with relative systematic uncertainty of $\pm5\%$), are corrected. The correlated yields are normalized by the number of trigger particles, $N_{\rm trig}$.

The two-particle acceptance correction is obtained from the mixed-events technique. The mixed events are required to be within 5~cm in $z_{\rm vtx}$, with the same multiplicity (for the TPC and FTPC-Au centrality measures) or within 10 attenuated ADC counts\footnote{
The ZDC-Au attenuated ADC dynamic range is 0-255 counts.} (for the ZDC-Au centrality measure). The mixed-events acceptance is normalized to 100\% at $\deta|_{100\%}=0$ for TPC-TPC correlations and at $\deta|_{100\%}=\pm3.3$ for TPC-FTPC correlations.

\begin{figure*}
\begin{center}
\includegraphics[width=0.4\textwidth]{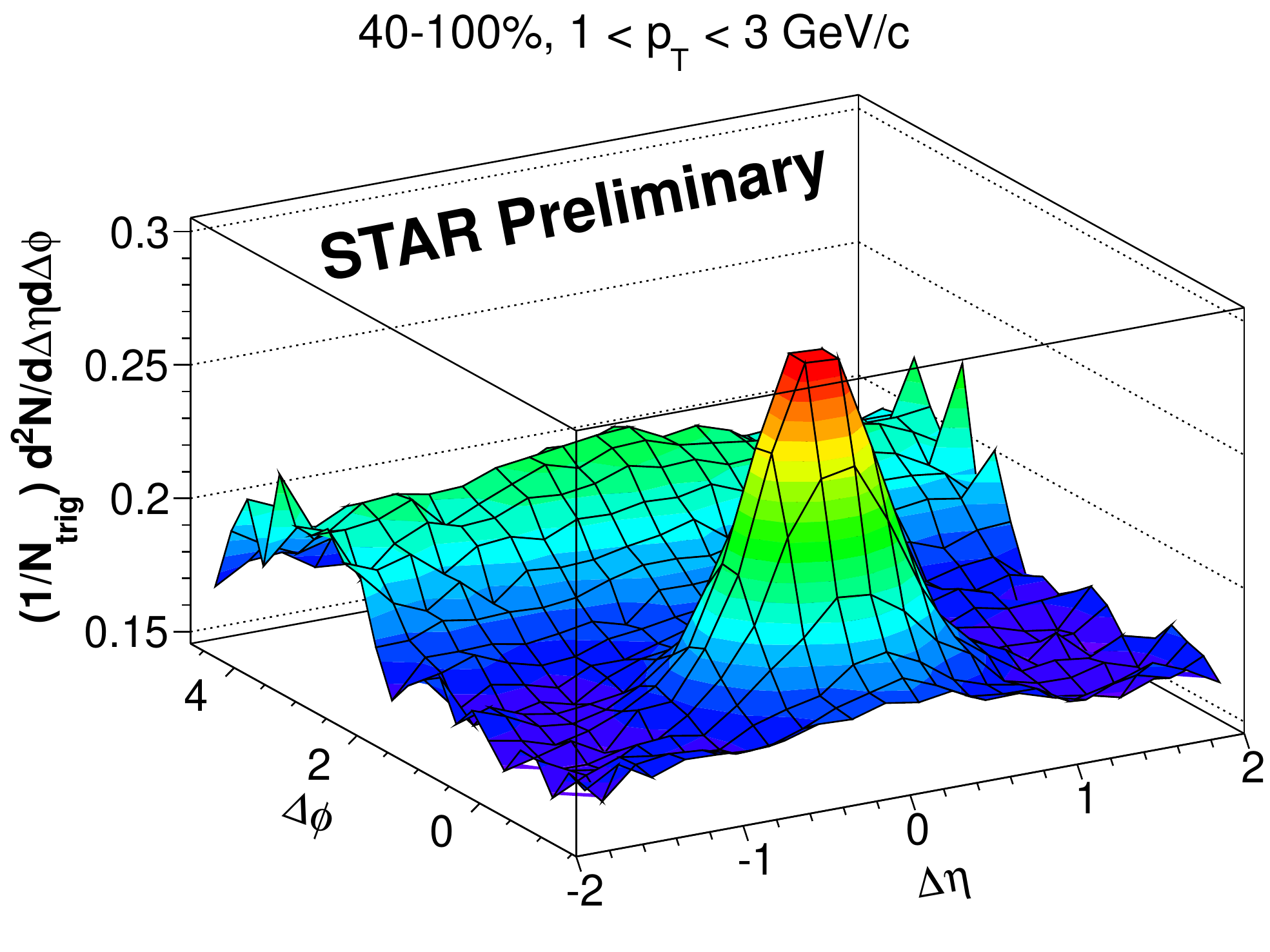}
\includegraphics[width=0.4\textwidth]{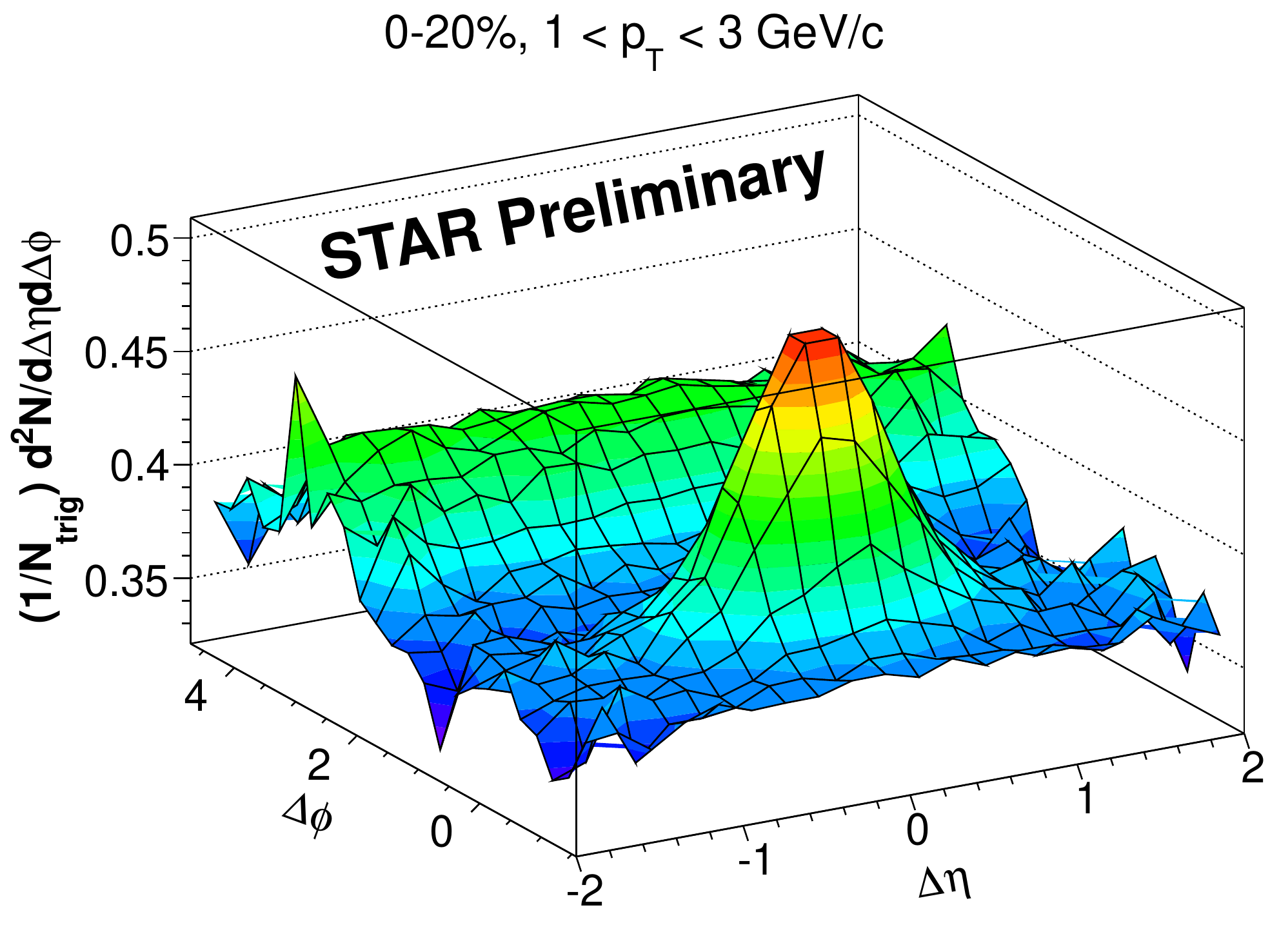}
\caption{Dihadron correlations in peripheral 40-100\% (left panel) and central 0-20\% (right panel) \dau\ collisions. Centrality is determined by FTPC-Au ($-3.8<\eta<-2.8$) multiplicity. Trigger and associated particle $\pt$ are both $1<\pt<3$~\gev. Note suppressed zeros.}
\label{fig:2d}
\end{center}
\end{figure*}

Figure~\ref{fig:2d} shows the two-particle acceptance corrected dihadron correlations in $(\deta=\eta_{\rm assoc}-\eta_{\rm trig},\dphi=\phi_{\rm assoc}-\phi_{\rm trig})$ between associated and trigger particle pseudo-rapidities and azimuthal angles, respectively, in peripheral and central \dau\ collisions. Namely,
\begin{equation}
\frac{1}{N_{\rm trig}}\frac{d^2N}{d\deta d\dphi}=\frac{1}{N_{\rm trig}}\frac{S(\deta,\dphi)/\epsilon_{\rm assoc}}{B(\deta,\dphi)/B(\deta|_{100\%},\dphi)}\,,
\end{equation}
where $S=\frac{1}{N_{\rm trig}}\frac{d^2N^{\rm same}}{d\deta d\dphi}$ is the raw pair density from same event and $B=\frac{1}{N_{\rm trig}}\frac{d^2N^{\rm mix}}{d\deta d\dphi}$ is the counterpart from mixed event. 
In the rest of this article, correlation functions projected onto $\dphi$ and $\deta$ are studied. Two approaches are taken to analyze the correlation functions. One is to analyze the correlated yields after subtracting a uniform combinatorial background. The background is normalized by the Zero-Yield-At-Minimum (ZYAM) assumption~\cite{ZYAM} in each $\deta$ bin. The other is to decompose the correlation functions into Fourier series and study the Fourier coefficients. No background subtraction is required; the interpretation of the Fourier coefficients, however, requires a physical model and possibly background consideration.

Systematic uncertainties are assessed by varying the ZYAM normalization $\dphi$ range from the default of 0.4 to 0.2 and 0.6 radian. In addition a 5\% systematic uncertainty from the tracking efficiency is applied on the correlated yield. For the Fourier coefficients, the systematic uncertainties are expected to be small compared to statistical uncertainties, but a thorough study of systematic uncertainties has not been done yet.

\section{Results on correlated yields}

\begin{figure*}
\begin{center}
\includegraphics[width=0.32\textwidth]{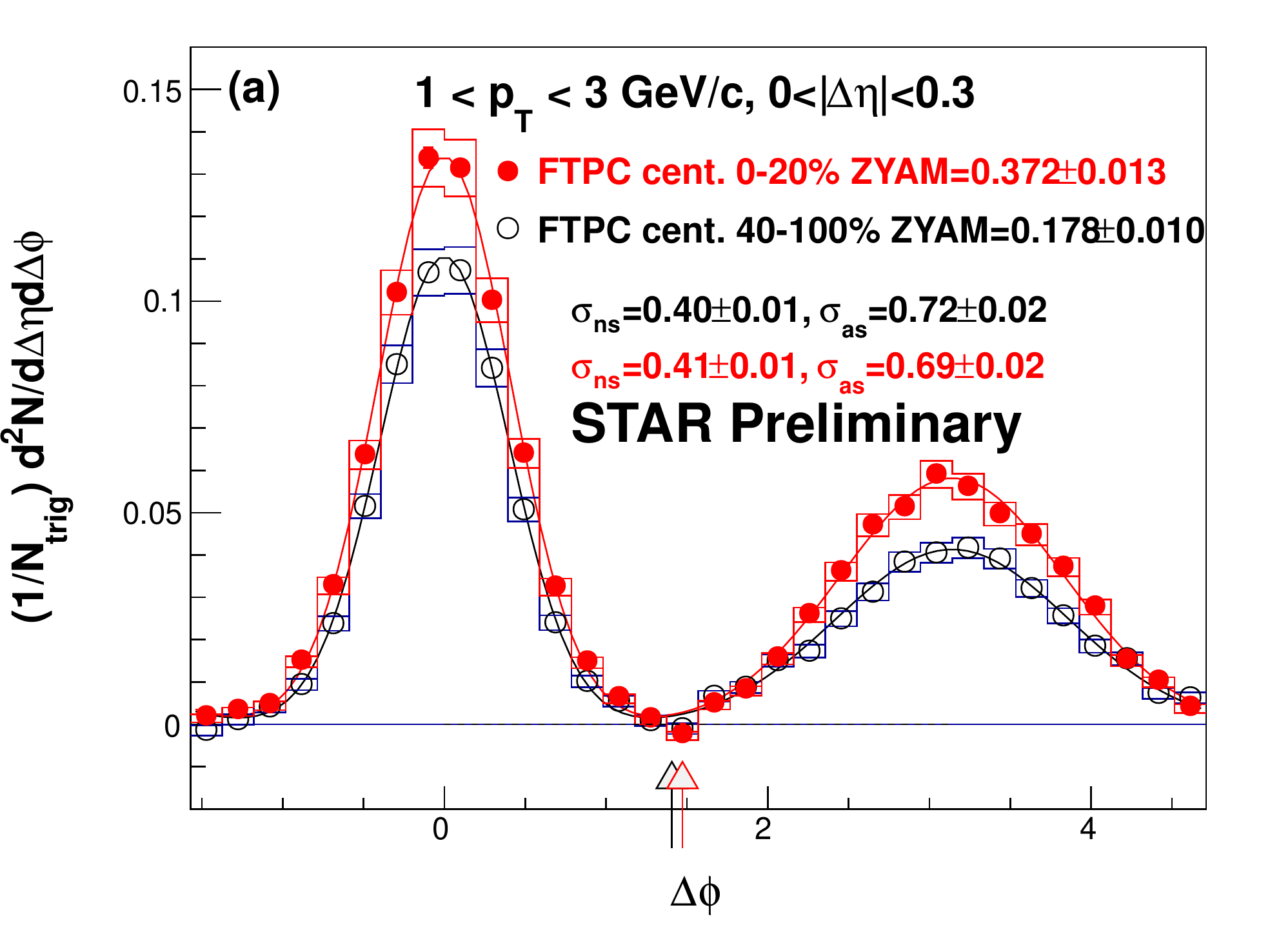}
\includegraphics[width=0.32\textwidth]{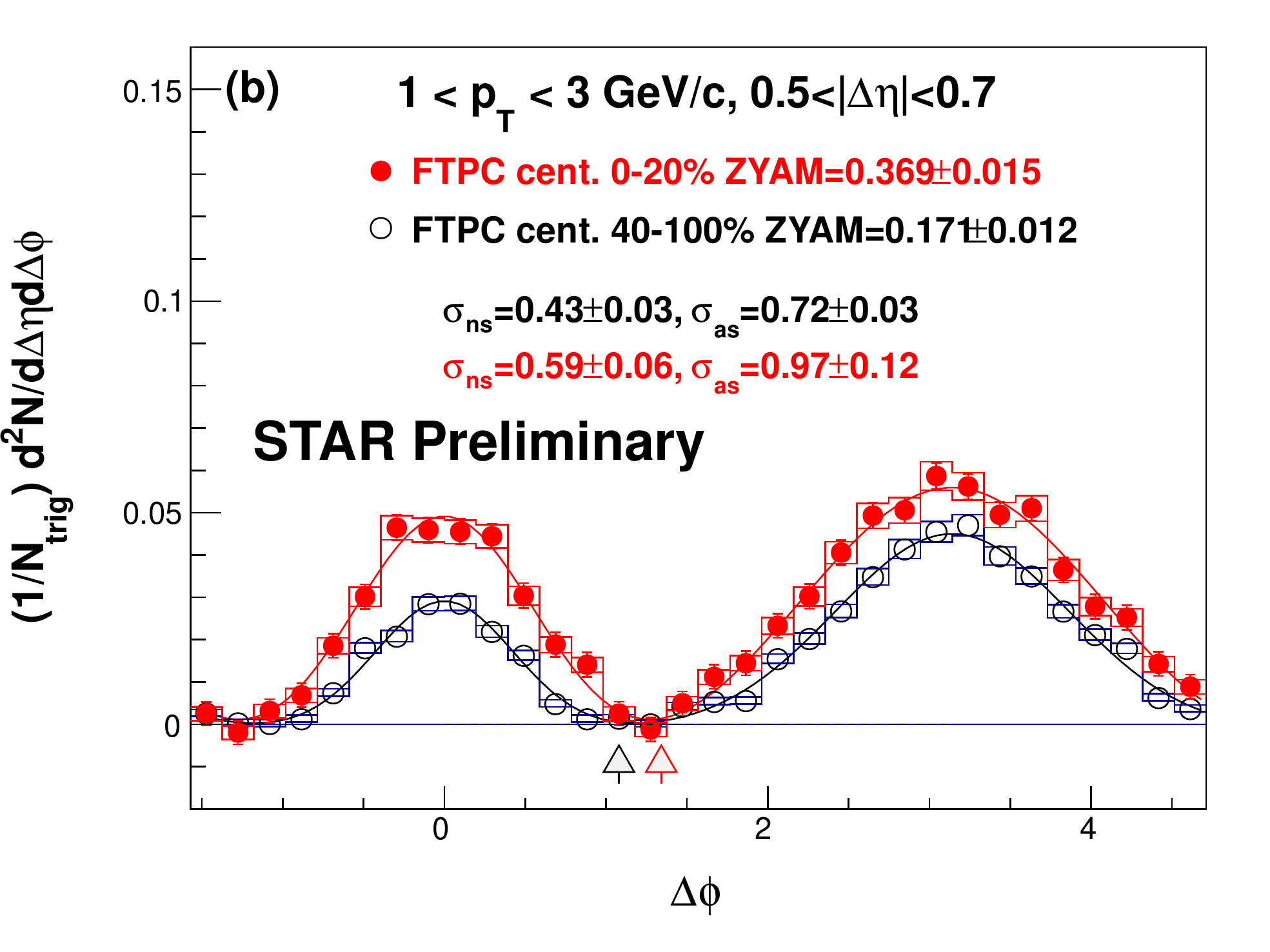}
\includegraphics[width=0.32\textwidth]{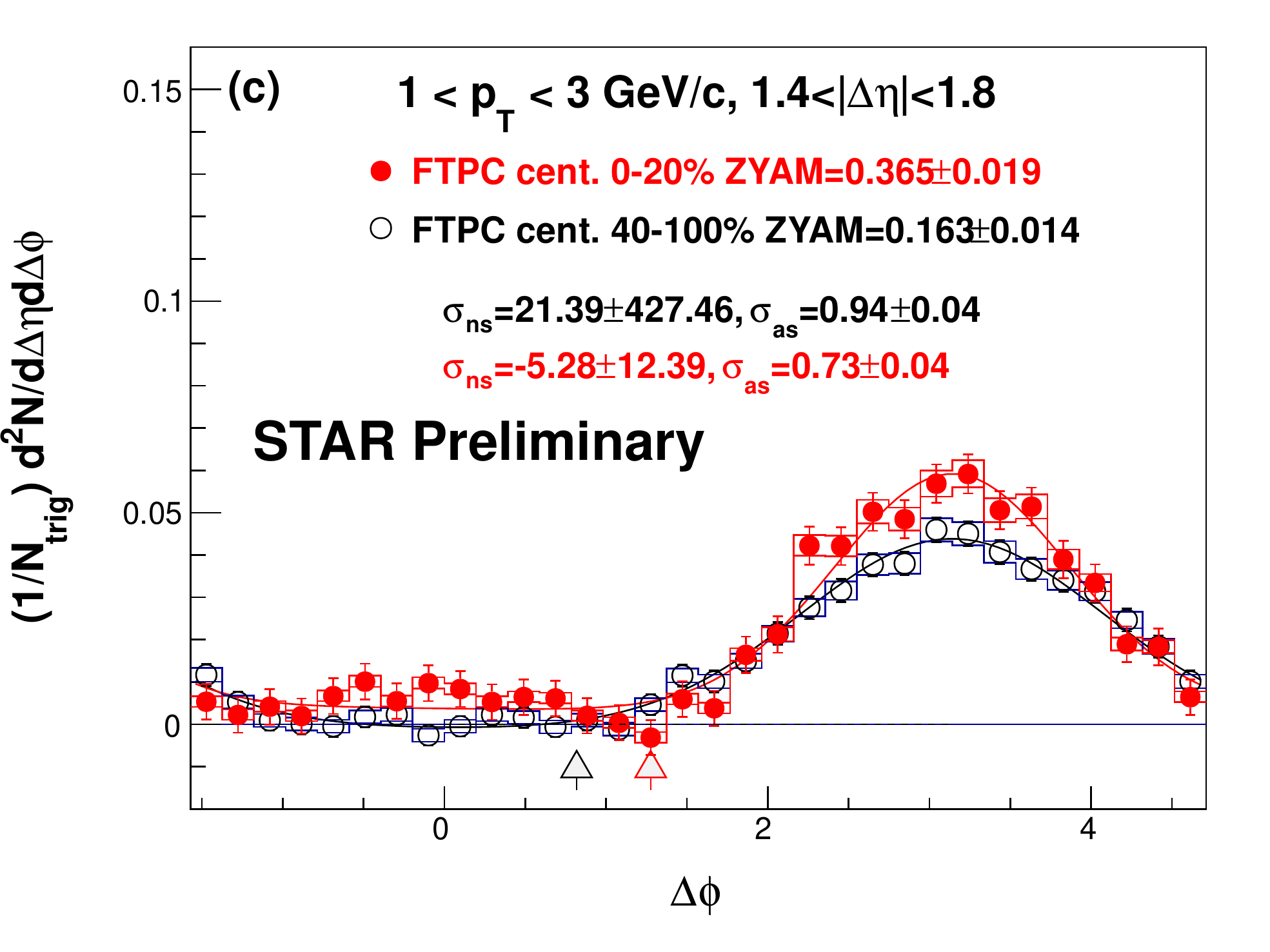}
\caption{Dihadron $\dphi$ correlations in three ranges of $|\deta|$ in peripheral (black) and central (red) \dau\ collisions. Trigger and associated particles are both from TPC ($|\eta|<1$) and $1<\pt<3$~\gev. Centrality is determined by FTPC-Au ($-3.8<\eta<-2.8$) multiplicity. The arrows indicate ZYAM normalization locations. Error bars are statistical and boxes indicate systematic uncertainties.}
\label{fig:dphi}
\end{center}
\end{figure*}

Figure~\ref{fig:dphi} shows the TPC-TPC $\dphi$ correlations in three ranges of $\deta$. Both peripheral and central collisions are shown; centrality is determined by the FTPC-Au. It is observed that the correlated yields are larger in central than peripheral \dau\ collisions. Difference between central and peripheral data will be shown in Fig.~\ref{fig:dphiDiff}.

\begin{figure*}
\begin{center}
\includegraphics[width=0.32\textwidth]{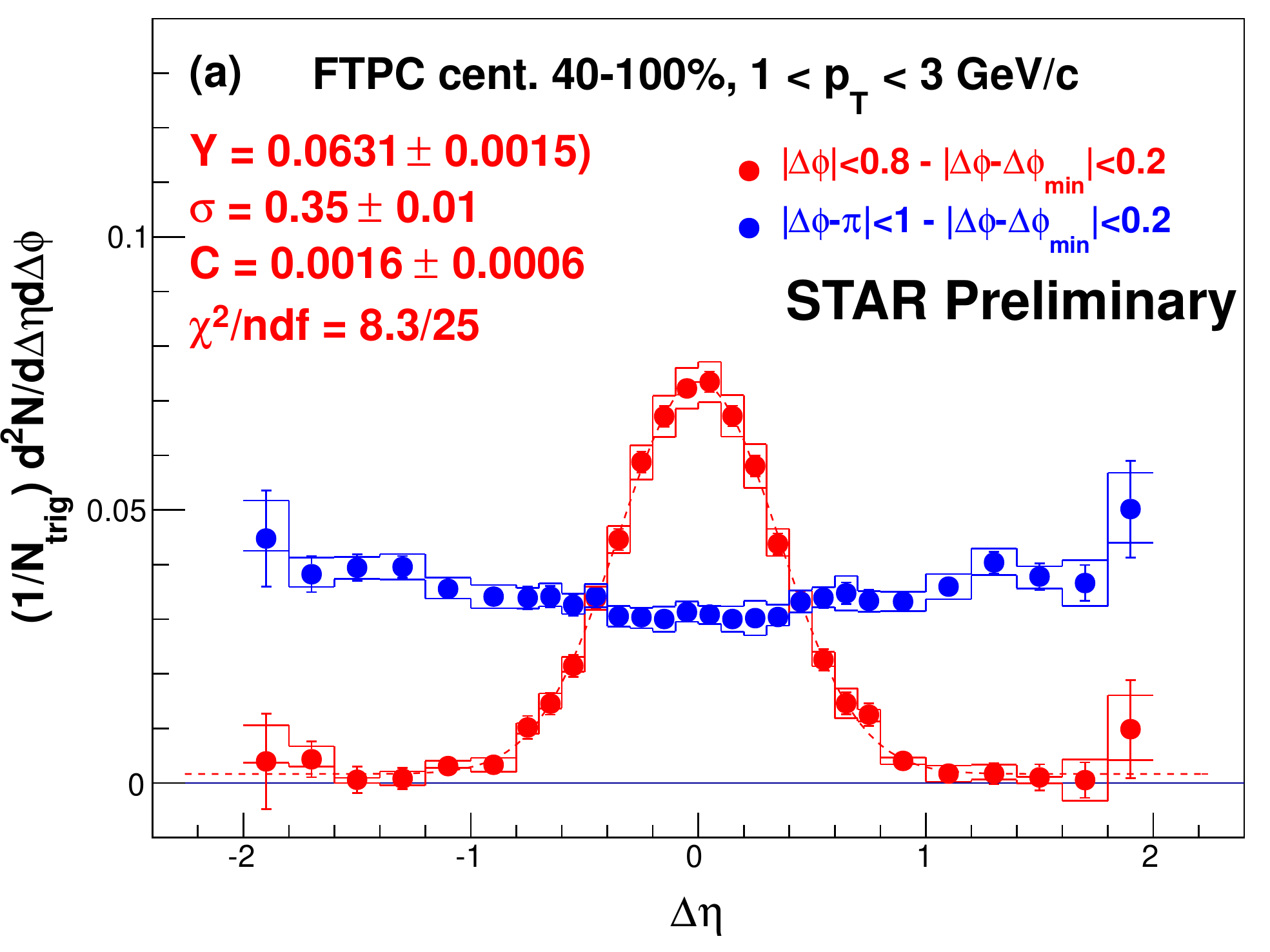}
\includegraphics[width=0.32\textwidth]{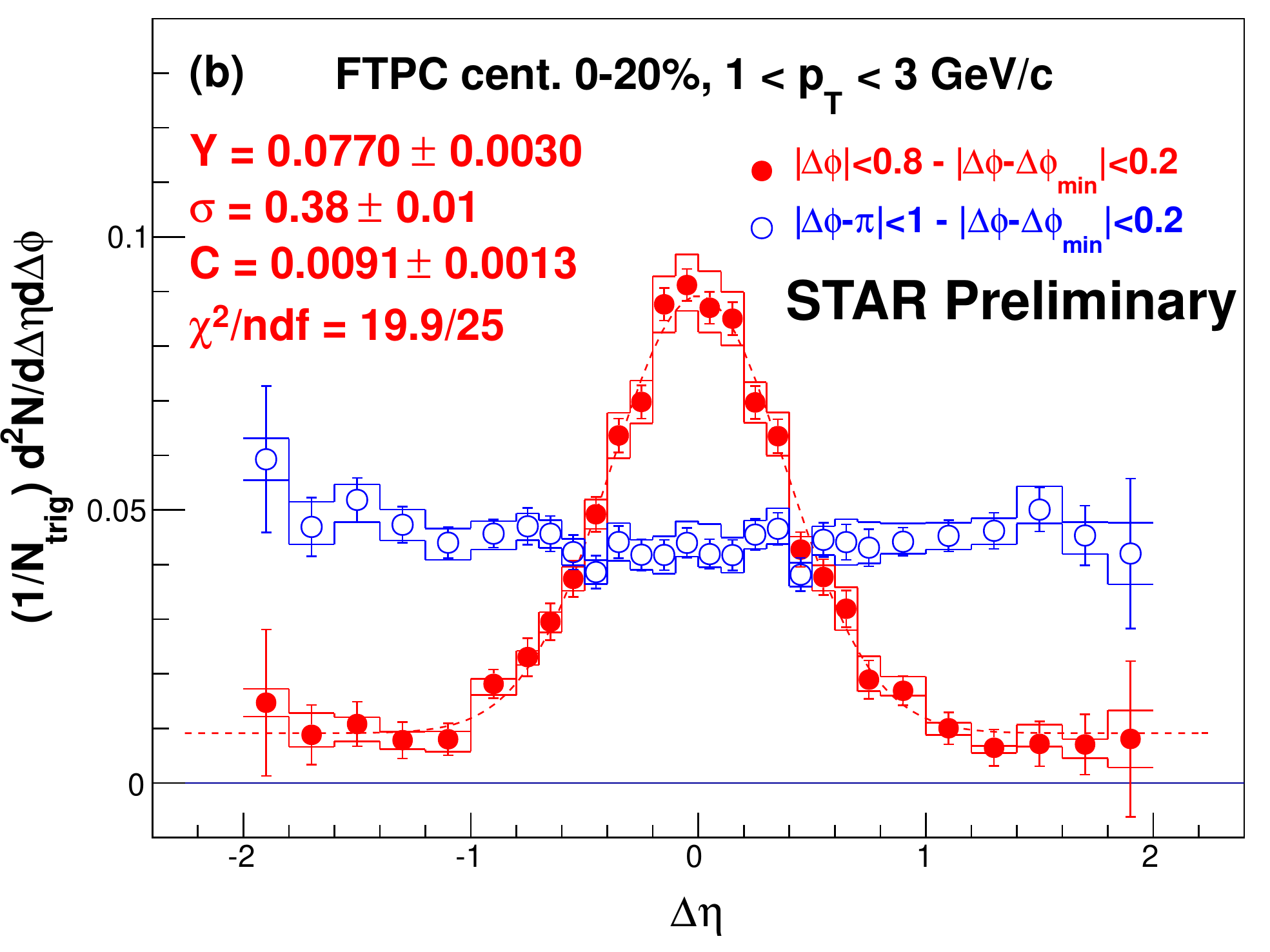}
\includegraphics[width=0.32\textwidth]{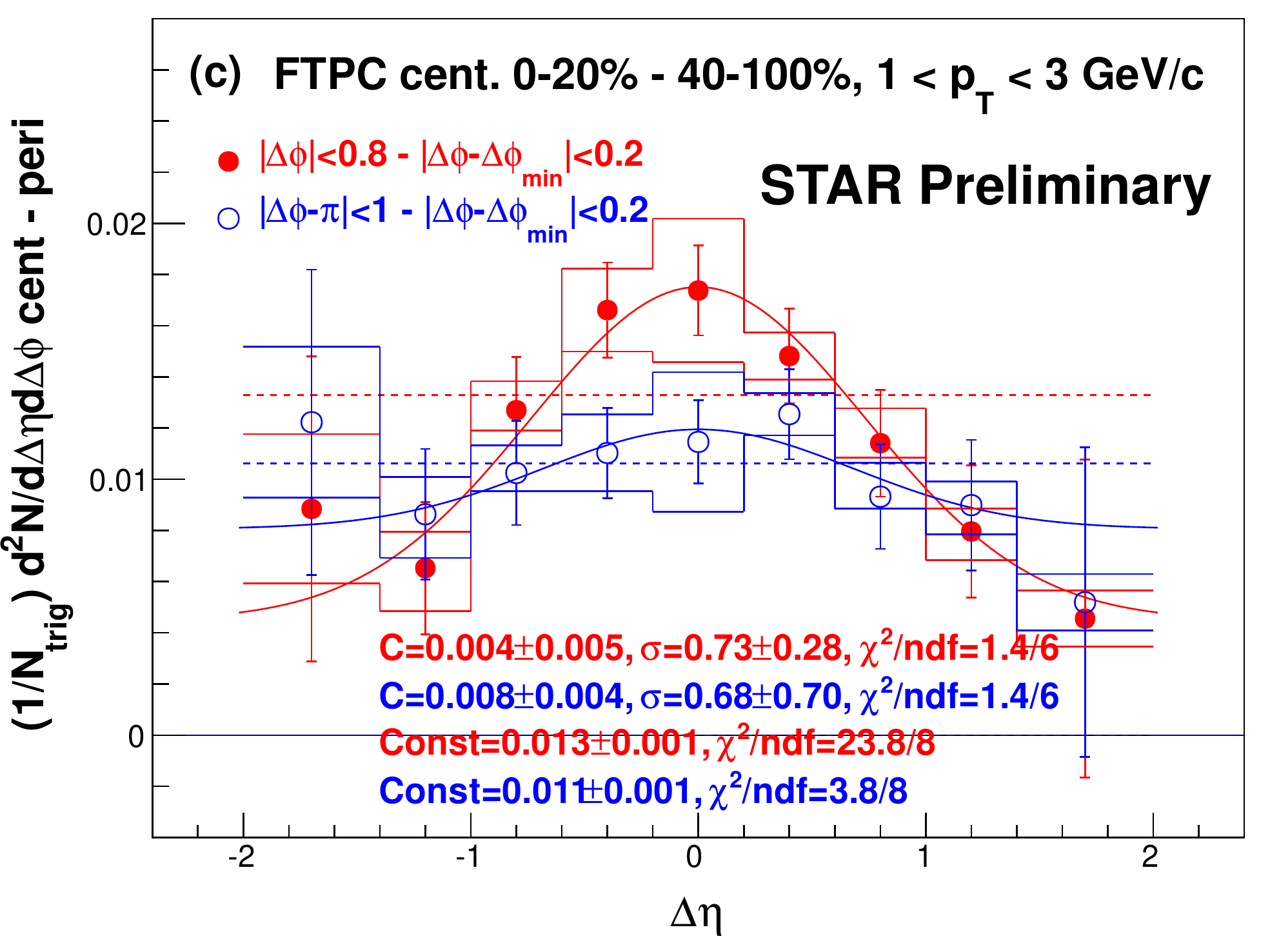}
\caption{Dihadron $\deta$ correlations for near side (red) and away side (blue) in peripheral (left panel) and central (middle panel) \dau\ collisions, and their ``central $-$ peripheral'' differences (right panel). Trigger and associated particles are both from TPC ($|\eta|<1$) and $1<\pt<3$~\gev. Centrality is determined by FTPC-Au ($-3.8<\eta<-2.8$) multiplicity. Error bars are statistical and boxes indicate systematic uncertainties.}
\label{fig:deta}
\end{center}
\end{figure*}

In order to investigate the source of the differences between central and peripheral collisions, $\deta$ correlations for near side ($|\dphi|<0.8$) and away side ($|\dphi-\pi|<1$) are shown in Fig.~\ref{fig:deta}. The near-side correlations exhibit a Gaussian peak and the away-side correlations are approximately uniform. Gaussian+constant fits to the near-side correlations indicate a difference of 20\% in the Gaussian area between central and peripheral collisions. The difference between central and peripheral collisions, shown in the right panel of Fig.~\ref{fig:deta}, exhibits a near-side Gaussian peak and an approximate uniform away-side. These resemble the jet-correlation features, suggesting that the ``central $-$ peripheral'' difference is mainly due to a difference in jet-like correlations. 
Given that little difference was observed between minimum-bias \pp\ and minimum-bias \dau\ data and significant differences appear between central and peripheral \dau\ data after event selections by centrality, we conclude that the
difference is most likely caused by biases in the centrality determination--although FTPC-Au is used for centrality which is 3 units away from the correlation measurement, away-side jet-correlations can still contribute to the overall multiplicity in FTPC-Au.

\section{Results on Fourier coefficients}

\begin{figure*}
\begin{center}
\includegraphics[width=0.32\textwidth]{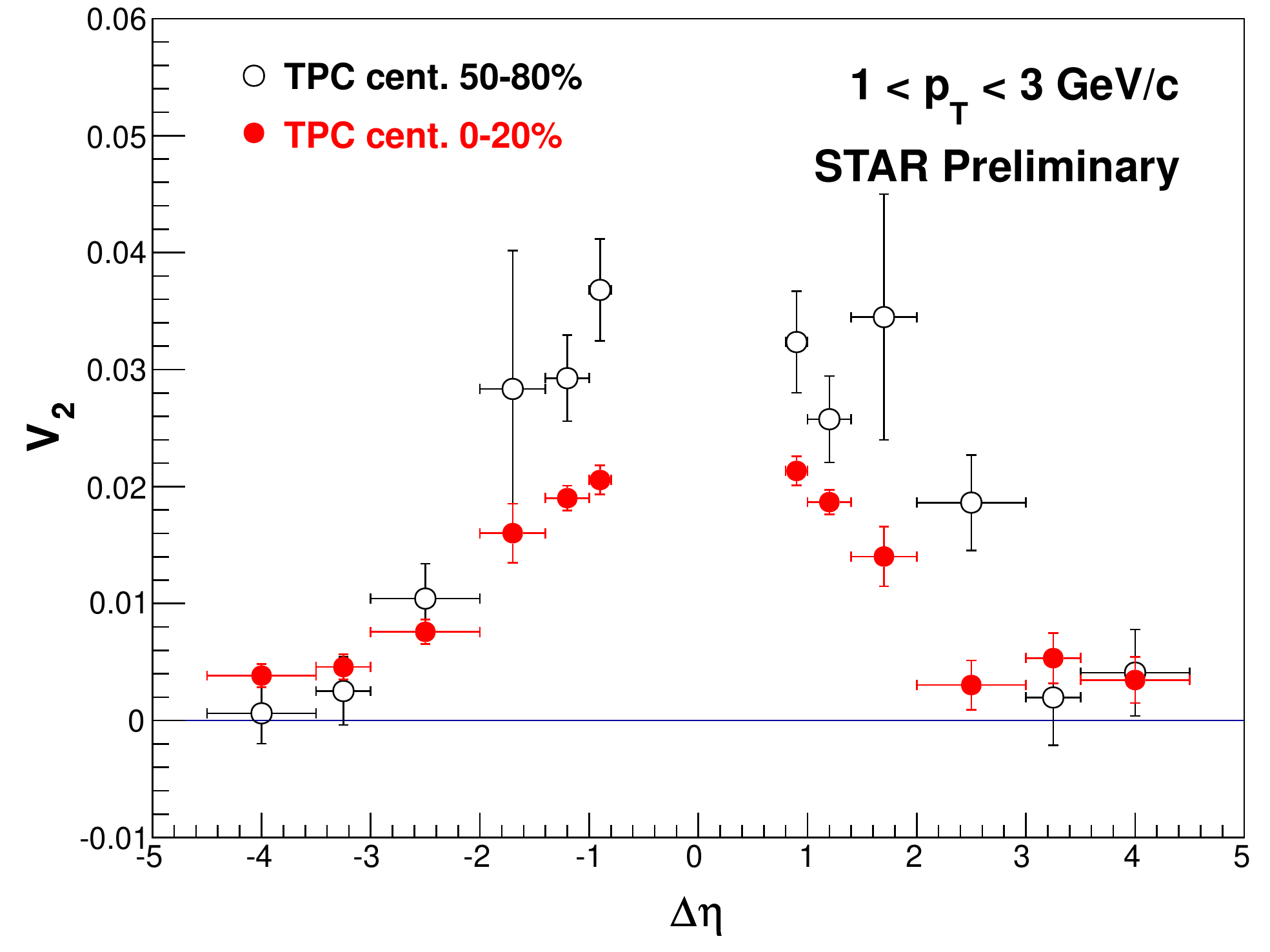}
\includegraphics[width=0.32\textwidth]{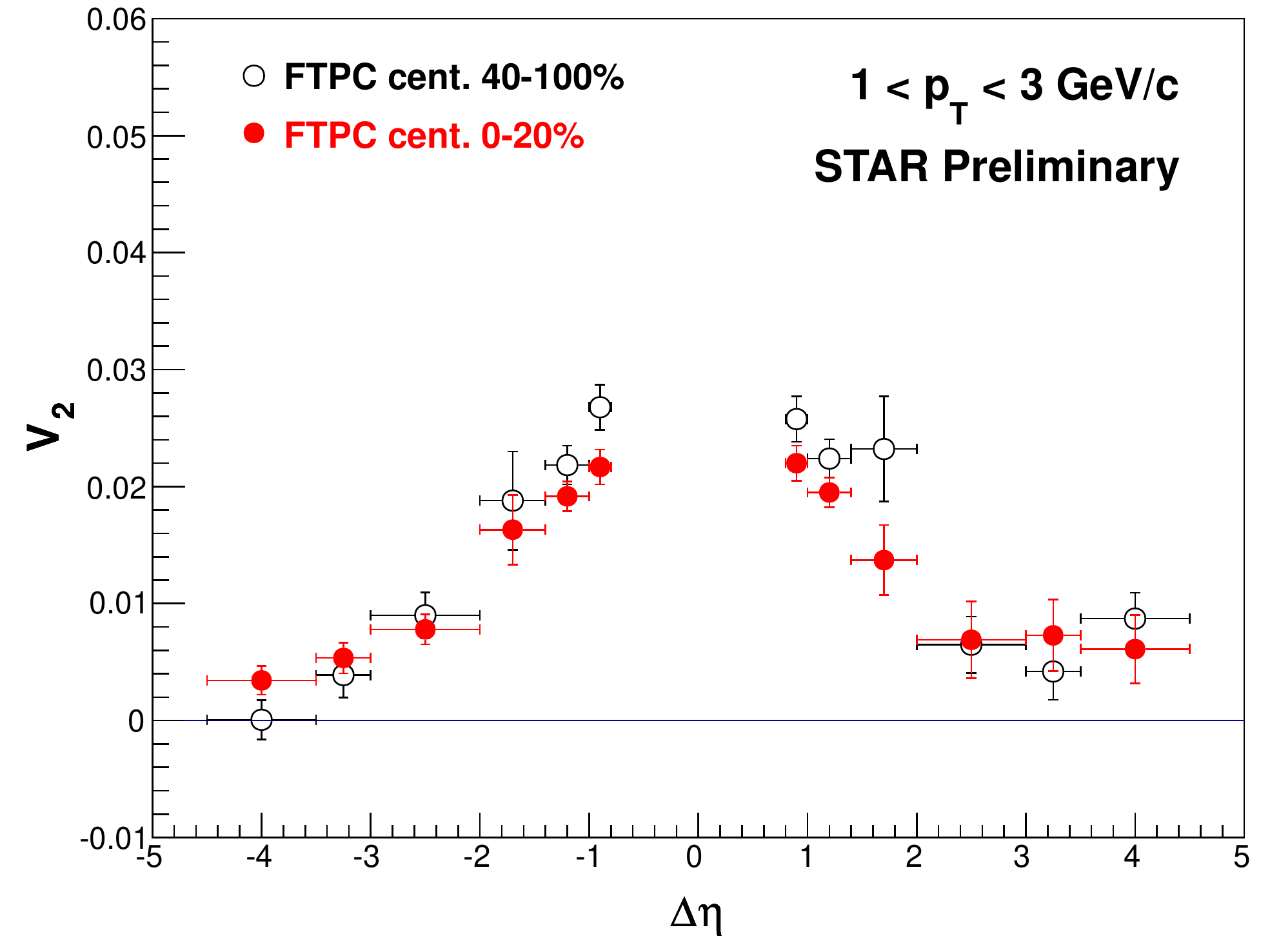}
\caption{Fourier coefficients of $\dphi$ correlation functions vs.~$\deta$ in peripheral (black) and central (red) \dau\ collisions. The data points within $-2<\deta<2$ are from TPC-TPC and the other data points are from TPC-FTPC correlations. Trigger and associated particle $\pt$ are both $1<\pt<3$~\gev. Centrality is determined by TPC multiplicity (left panel) and FTPC-Au multiplicity (right panel). Error bars are statistical.}
\label{fig:VnVsDeta}
\end{center}
\end{figure*}

Figure~\ref{fig:VnVsDeta} shows the second harmonic Fourier coefficients of the $\dphi$ distributions of the associated particle yields, $V_n=\mean{\cos n\dphi}$ $(n=2)$ where the average is taken over all histogram bins of the $\dphi$ distribution, as a function of $\deta$ for both peripheral and central collisions. The third harmonic Fourier coefficient ($n=3$) is consistent with zero. The centrality is determined by TPC in the left panel and FTPC-Au in the right panel; thus in the left panel the data points above $|\deta|>2$ (from TPC-FTPC correlations) are more relevant because they are less biased by the centrality measure, and in the right panel those at $|\deta|<2$ (from TPC-TPC correlations) are more relevant. The Fourier coefficients decrease with increasing $|\deta|$. This is consistent with a jet-like contribution to be primarily responsible for the measured $V_2$.

\begin{figure*}
\begin{center}
\includegraphics[width=0.32\textwidth]{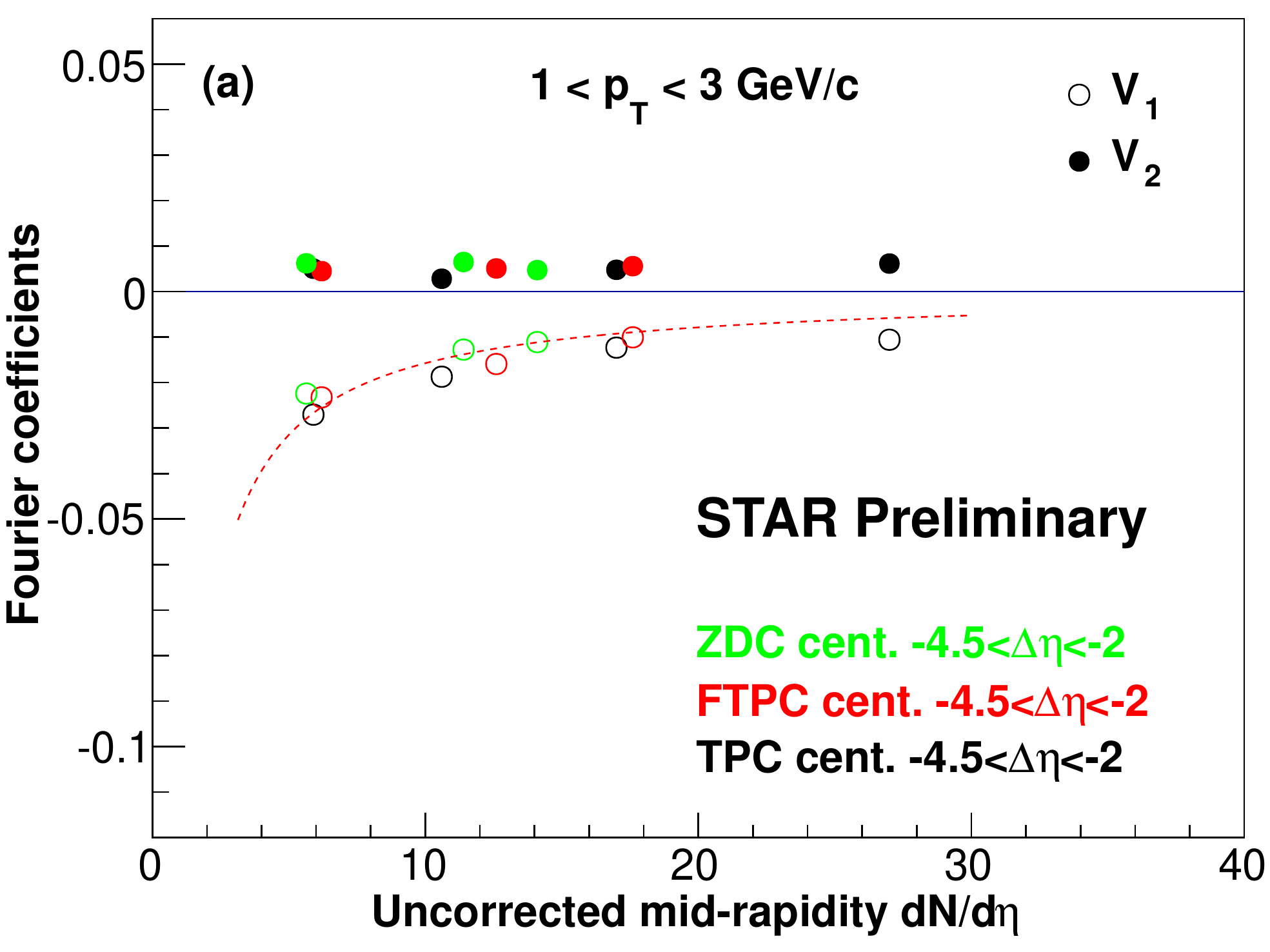}
\includegraphics[width=0.32\textwidth]{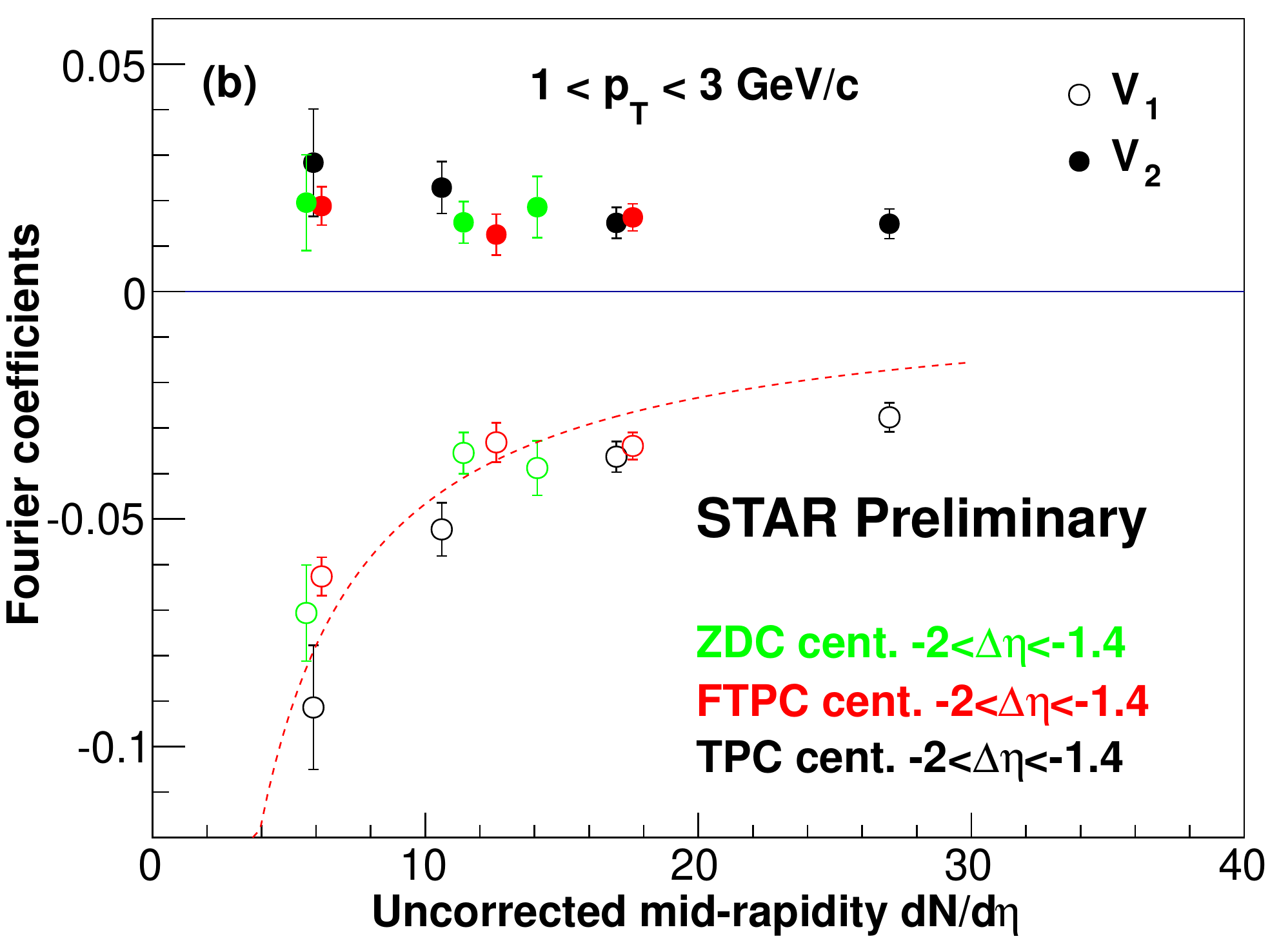}
\includegraphics[width=0.32\textwidth]{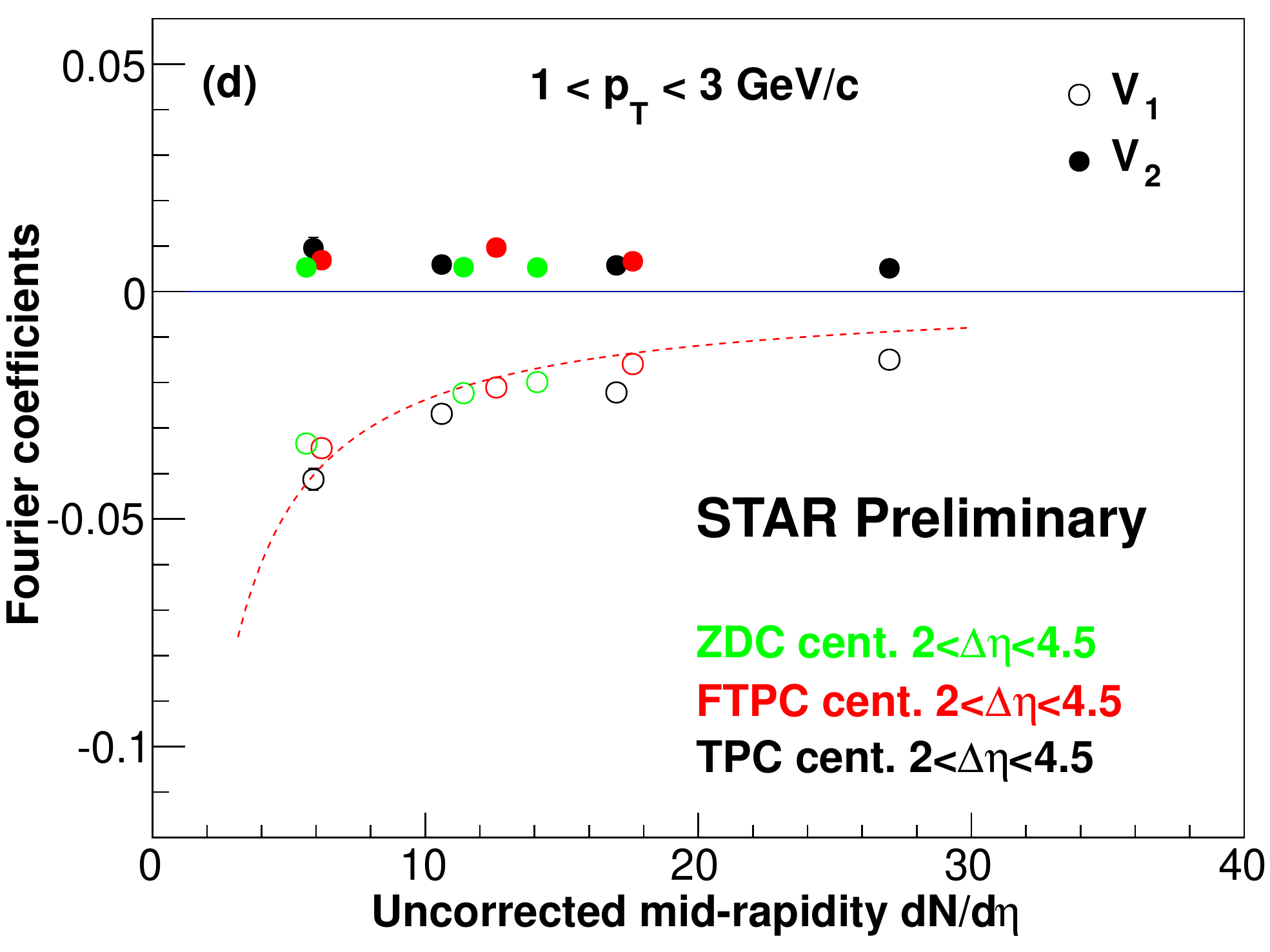}
\caption{Fourier coefficients of $\dphi$ correlation functions vs. event multiplicity in \dau\ collisions for three ranges of $\deta$ (as indicated in legends). Results from all three centrality definitions are shown. Trigger and associated particle $\pt$ are both $1<\pt<3$~\gev. Error bars are statistical.}
\label{fig:VnVsMult}
\end{center}
\end{figure*}

Figure~\ref{fig:VnVsMult} shows the Fourier coefficients, $V_1$ and $V_2$. ($V_3$ is consistent with zero.) 
Three ranges of $\deta$ are shown, from left to right, for TPC-FTPC-Au, TPC-TPC with negative $\deta$ (the TPC-TPC positive $\deta$ results are similar), 
and TPC-FTPC-d correlations. 
Results with all three centrality determinations are shown, plotted at the corresponding measured mid-rapidity charged particle multiplicity density $dN/d\eta$. The $V_1$ is observed to approximately vary as $(dN/d\eta)^{-1}$, while the $V_2$ is approximately independent of $dN/d\eta$. $V_2$ is finite at all measured $\deta$; it is larger at mid-rapidity than forward/backward rapidities; $V_2$ from TPC-FTPC-d correlation may be even larger than that from TPC-FTPC-Au correlation.

\begin{figure*}
\begin{center}
\includegraphics[width=0.32\textwidth]{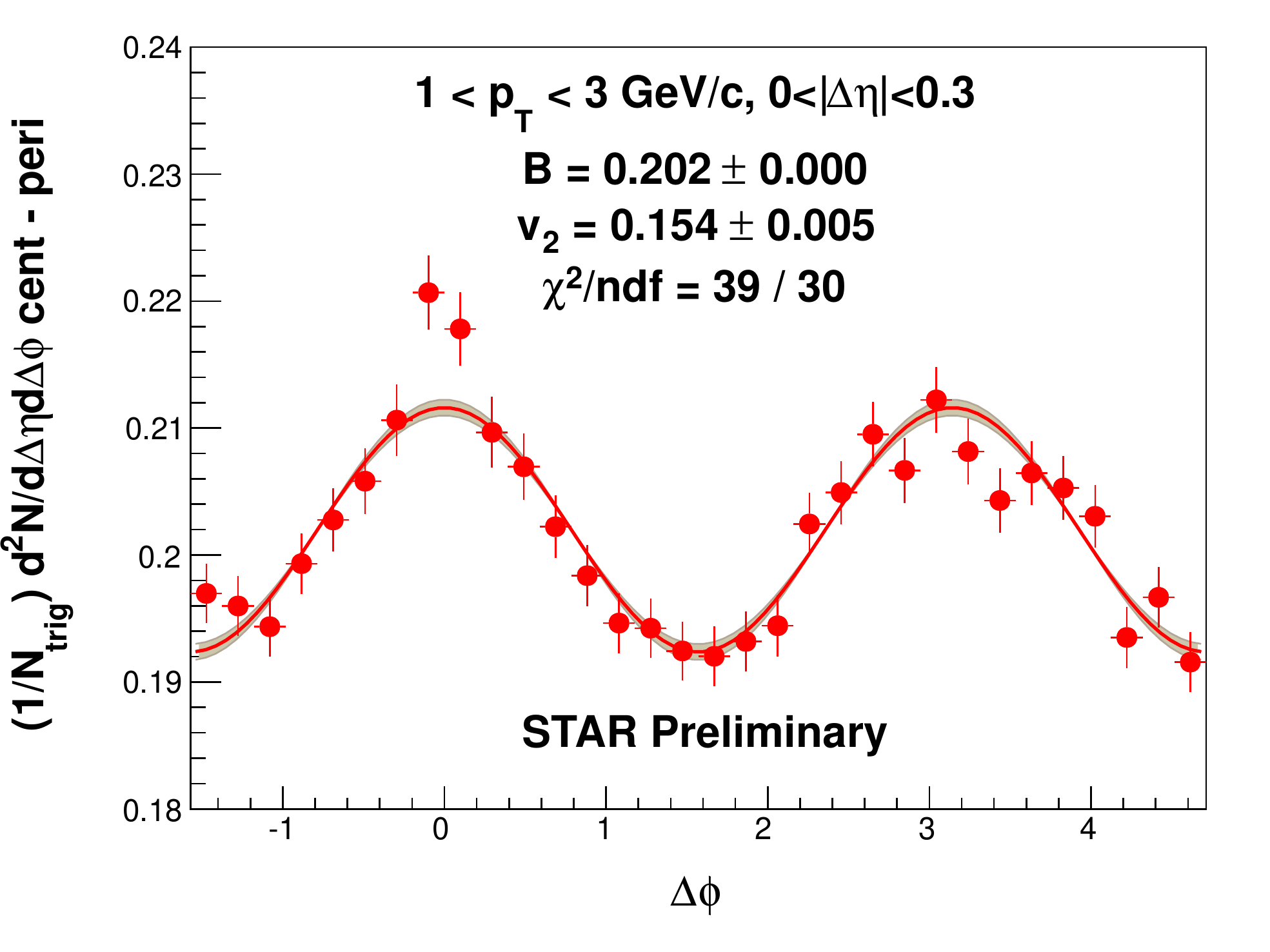}
\includegraphics[width=0.32\textwidth]{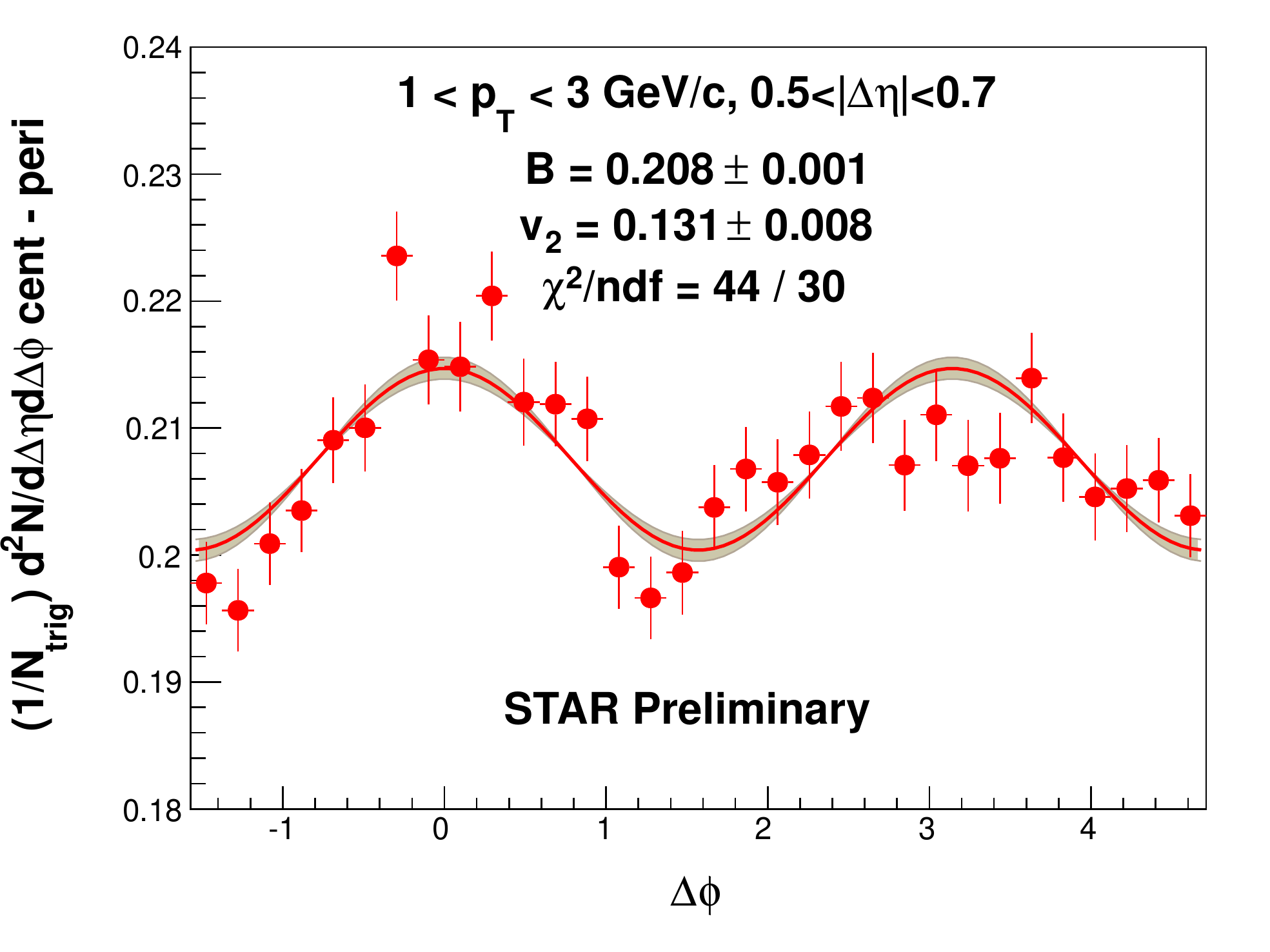}
\includegraphics[width=0.32\textwidth]{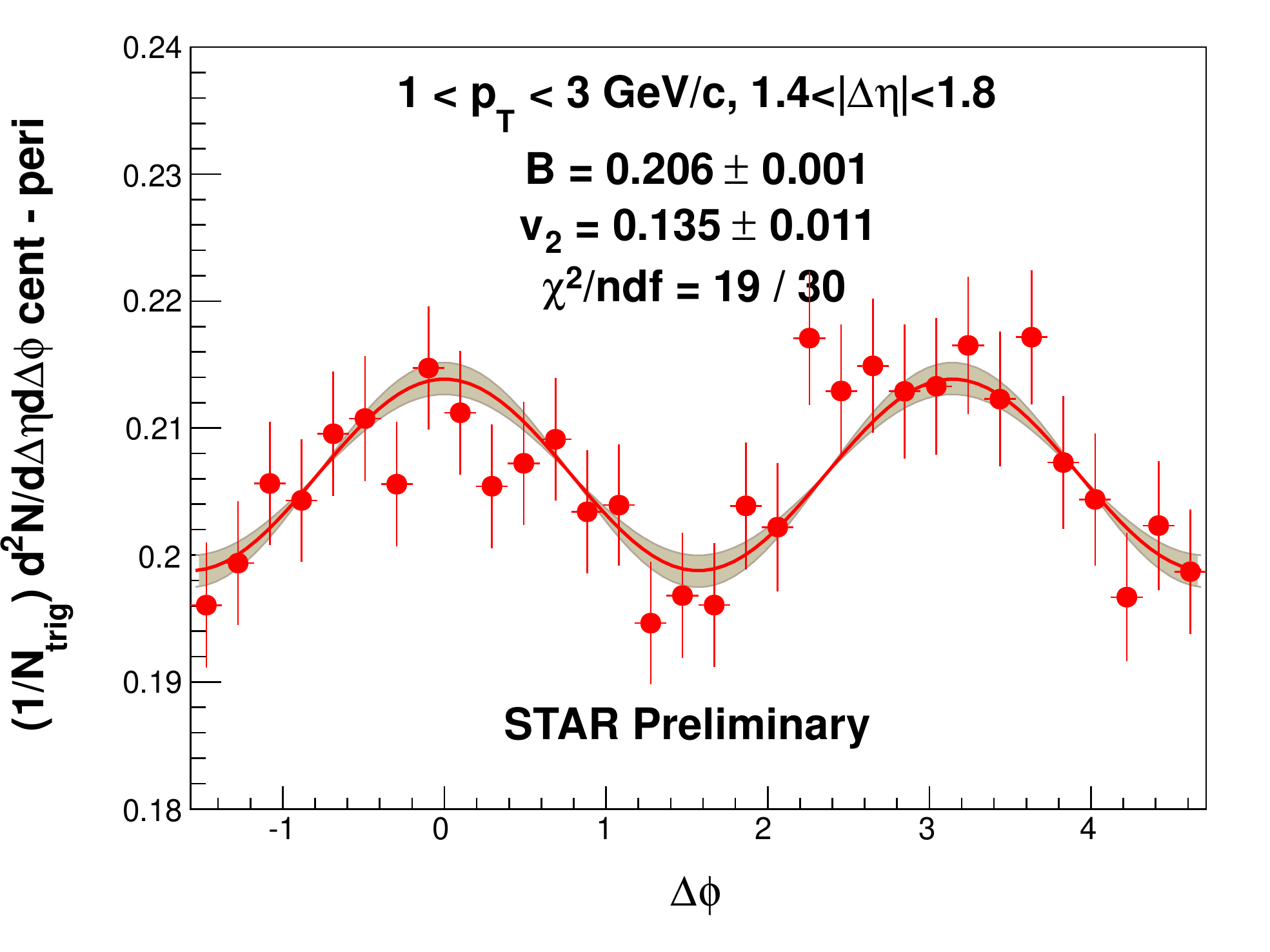}
\caption{Difference between dihadron $\dphi$ raw correlations in central and peripheral collisions, corresponding to those in Fig.~\ref{fig:dphi}.}
\label{fig:dphiDiff}
\end{center}
\end{figure*}

In fact, the Fourier coefficients of the ``central $-$ peripheral'' correlations are no different from those of the peripheral and central collisions. See Fig.~\ref{fig:dphiDiff} which shows the difference between the raw correlation functions in central and peripheral collisions corresponding to those in Fig.~\ref{fig:dphi}. 
This can be expressed in math as:
``central $-$ peripheral'' $=N_{\rm cent}(1+2V_1^{\rm cent}\cos\dphi+2V_2^{\rm cent}\cos2\dphi)-N_{\rm peri}(1+2V_1^{\rm peri}\cos\dphi+2V_2^{\rm peri}\cos2\dphi)\approx(N_{\rm cent}-N_{\rm peri})(1+2V_2\cos2\dphi)$ where $N_{\rm cent}V_1^{\rm cent}\approx N_{\rm peri}V_1^{\rm peri}$ ($N_{\rm cent}$ and $N_{\rm peri}$ are the numbers of associated particles in central and peripheral collisions, respectively) and $V_2\approx V_2^{\rm cent}\approx V_2^{\rm peri}$. However, the underlying physics mechanisms for the large Fourier coefficients of the ``central $-$ peripheral'' correlations are not entirely clear. Whether there are additional sources, except the aforementioned difference in jet-like correlations due to centrality biases, remains an open question. One of the future studies is to better quantify the centrality biases to jet-like correlations, and then investigate any additional physics mechanisms for the ``central $-$ peripheral'' difference. 

\section{Summary}

Dihadron $\dphi$ and $\deta$ correlations are reported for peripheral and central \dau\ collisions at $\snn=200$~GeV from STAR. The ZYAM background-subtracted correlated yields are larger in central than peripheral collisions. The ``central $-$ peripheral'' differences resemble jet-like correlations, Gaussian peaked on the near side and approximately uniform on the away side. The difference is mainly caused by difference in jet-like correlations due to centrality biases. Fourier coefficients of the raw dihadron correlations are also reported. The first harmonic coefficient is found to be approximately inversely proportional to event multiplicity. The second harmonic coefficient is found to decrease with $\deta$, but finite at forward/backward rapidity of $|\deta|\approx3$; it is approximately independent of the event multiplicity. 

The large acceptance of STAR allows detailed investigation of dihadron correlations and their centrality biases. The \dau\ data seem to be mainly consistent with jet phenomenology. The next step is to quantify ``central $-$ peripheral'' differences caused by centrality biases, and hopefully isolate possible additional contributions unrelated to jets.

\section*{Acknowledgments}
The author thanks Dr.~Carlos Salgado and the other organizers of IS-2013 for invitation and the stimulating conference. This work was supported by U.S.~Department of Energy under Grant No. DE-FG02-88ER40412.


\end{document}